\documentclass[a4paper,11pt]{article}
\pdfoutput=1 

\usepackage{jinstpub} 

\title{Developing an accurate and robust tool for pixel module characterization}

\author[a,1,2]{D. Hohov,\note{Corresponding author.}\note{Now at Université libre de Bruxelles, Belgium.}}
\author[a,3]{A. Lounis, \note{Also at Université Paris-Saclay, France.}}
\author[a]{A. Falou,}
\author[b]{E. - L. Gkougkousis}
\author[a,4]{and A. Bassalat \note{Also at An Najah National University, Nablus, Palestine.}}

\affiliation[a]{Laboratoire de physique des deux infinis Irène Joliot-Curie (IJCLab), CNRS-IN2P3, Université Paris-Saclay,\\Bâtiment 100, 15 rue Georges Clémenceau, 91405 Orsay, France}
\affiliation[b]{CERN,\\Espl. des Particules 1, 1211 Meyrin, Switzerland}

\emailAdd{dmytro.hohov@cern.ch}

\abstract{For operation at the High Luminosity Large Hadron Collider (LHC), the ATLAS experiment is building a new all-silicon inner tracker (ITk). The production and testing of thousands of silicon pixel and strip modules is required to cover the estimated 180 m$^{2}$ of the total surface area. A compact, affordable and robust module characterization system is required for in-situ testing prior any test beam campaign or installation. A test bench setup based on an infrared laser was developed at IJCLab, allowing also for the use of micro-metric precision scanning with a radioactive source. A detailed schema of the setup, operating principles and testing methods are described in this paper, together with the first results obtained with a FE-I4 silicon pixel module.}


\arxivnumber{2112.07633} 

\begin{document}
\maketitle
\flushbottom
\section{Introduction}

\label{sec:intro}

In the context of the High Luminosity phase of LHC (HL-LHC) envisaged to start in 2026, it is planned to increase the instantaneous beam luminosity up to 7.5$\times$10$^{34} cm^{-2}s^{-1}$ to deliver a total integrated luminosity of 4000 fb$^{-1}$~\cite{tdr}. Such operating conditions lead to additional requirements for the detector systems for the HL-LHC phase, since along with the increase of luminosity, the workload and radiation degradation of the detectors becomes also higher. The current ATLAS Inner Detector (ID) cannot withstand the new harsher environment, where the resulting particle flux and radiation levels are expected to exceed design values by one order of magnitude up to the maximum 1 MeV neutron equivalent fluence of 1.31$\times$10$^{16} cm^{-2}$ for the first pixel layer~\cite{tdr}. As a result, a complete replacement of the entire ID is planned with a new all-silicon tracker system, called the Inner Tracker (ITk). It consists of a pixel detector close to the beam line and a large-area strip tracking detector surrounding it~\cite{tdr}. The ITk will use advanced sensors and readout electronics technologies providing the same or, in most cases, better performance than the existing ID in an environment with significantly higher pile-up and radiation levels~\cite{b}. Considering this fact, the new high granularity silicon sensors of enhanced geometrical efficiency with the capability to withstand enormous radiation damage are developed and have to be produced.

\subsection{Characterization of silicon pixel detectors}
An important step in new silicon pixel detector modules R\&D is testing and characterization in order to determine their performance and efficiency in a similar environment to the one expected at the HL-LHC. After the module assembly and wire-bonding, the properties of the detector prototypes have to be tested and evaluated, at first, performing IV measurements, readout chip tunings, temperature hardness tests. As a final step prior to the detector large scale module production and installation in the experiment, the performance of the modules is evaluated on a high energy particle beam. However, it is not always convenient to use relativistic particle beams for module tests due to limited availability and accessibility of the relevant facilities. In addition, during such campaigns, a complicated and accurate reconstruction procedure is necessary to establish the hit map of particles crossing the tested module.
Several alternative options are nevertheless available for efficient module testing under realistic conditions. In such implementations, one needs to reproduce in the laboratory the amount of charge generated by a MIP (minimum ionizing particle) in the sensor volume.

One possibility would be to use cosmic rays or $\beta$-radioactive sources, for example 90-Sr, which emits electrons of energy $E_{max}=2.28 MeV$~\cite{rad}, high enough to behave as a MIP. The drawback of such an approach is the energy of the $\beta$-spectrum is broad and low energy particle deposit more energy than MIPs along the trajectory. Furthermore, the exact position of interaction with the sensor cannot be determined. Additionally, the particle rate of a small laboratory radioactive source or cosmic rays is significantly low compared to the accelerated particle beams.

Another alternative would be the usage of a laser beam to deposit charge within the silicon sensor volume comparable to a MIP crossing. The latter will be discussed in detail in the next section.

\section{Pixel module characterization using a laser beam}
A laser oriented system allows for flexible charge injection at a well-defined position and depth within the silicon die due to the well-known dimensions and energy of the laser beam. The injection depth can be adjusted by varying the laser light wavelength and hence the corresponding penetration depth. The amount of deposited charge can be set by modifying the laser intensity. Since the periodicity of the pulse can be set to kHz or MHz frequency, it is possible to achieve a desired amount of charge injections in a very short time.

Due to the high spatial precision of the detector module positioning with the X-Y table and the relatively small laser spot size below 20 µm, it is feasible to study effects, appearing at the edge of a sensor with high accuracy and statistics. At the same time, the focusing of the laser beam to the small spot size close to diffraction limits, makes the system more challenging for implementation. In addition, the effect of reflection from an aluminum surface and oxide-Si boundaries as well as the beam refraction makes it difficult to calculate precisely the amount and the position of injected charge. Moreover, the laser light interaction mechanism is different from the particle interaction, for example, the charge generated by a laser beam is almost constant, while for a MIP it follows the Landau distribution, so the effects should be carefully studied to obtain the reliable results.

\subsection{Description of the principles}
Typically, infrared lasers are used, due to the specific light interaction with silicon in this wavelength range. The energy of the photon with a wavelength $\lambda$ = 1060 nm is 1.17 eV, which is slightly above the energy gap of silicon at room temperature ( $E_g$ = 1.12 eV). It makes possible the photon to be absorbed by photoelectric effect. Moreover, the intensity of the light penetrating through silicon drops exponentially. The range for the 635 nm photon is 2-3 µm, whereas at 1060 nm it is about 850 µm in depth into the silicon material~\cite{dd}. This means that a pulse of well collimated infrared laser light of this wavelength can easily traverse 200 µm thick silicon sensor bulk, producing a column of uniformly distributed charge, to simulate MIP charge generation.

The setup can use two laser sources with different laser wavelengths (red and infrared range), making it possible to inject charge only on the surface of the sensor or throughout the whole its thickness. The red laser can also be used for the alignment since, unlike the infrared, it is visible by the human eye.

\begin{figure}[htbp]
\centering 
\includegraphics[width=.45\textwidth]{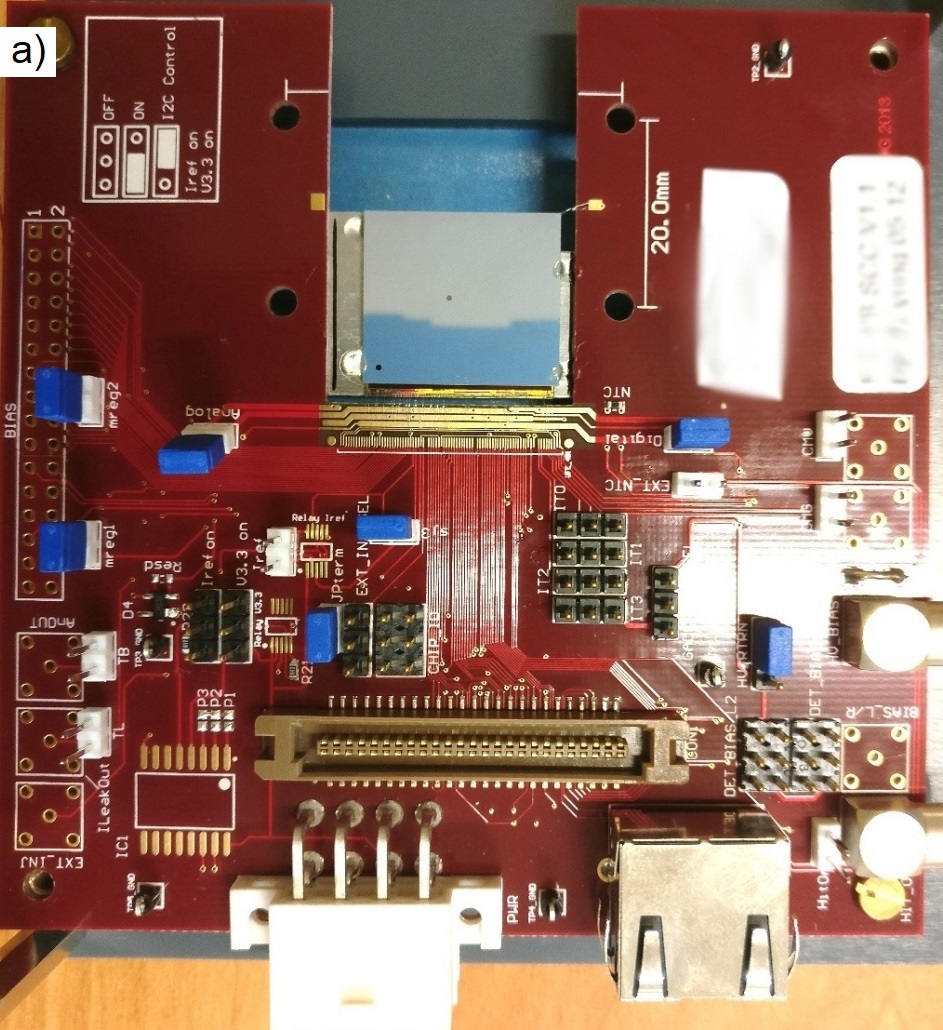}
\includegraphics[width=.45\textwidth]{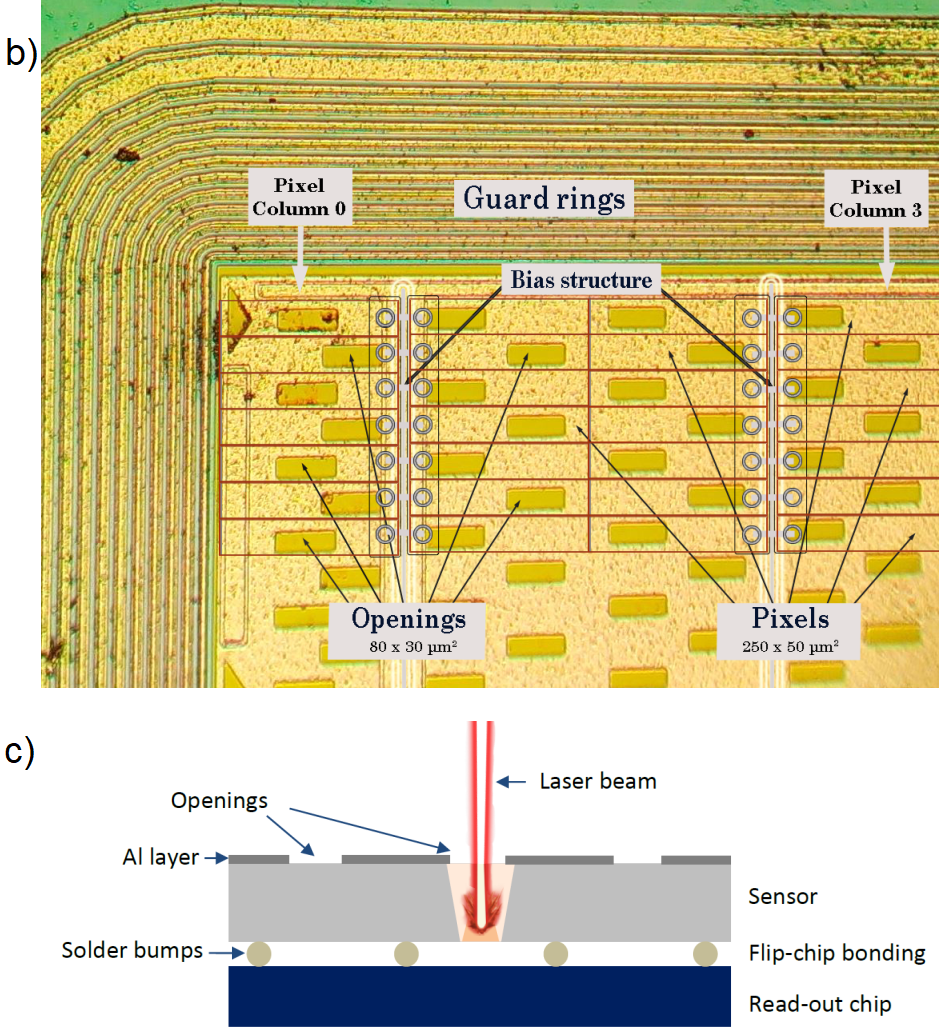}
\caption{\label{fig:SCC}a) A picture of a single chip card with a FE-I4 module. b) The backside view of IBL pixel sensor design with openings. c) A schema of laser tests with a pixel sensor module.}
\end{figure}

For the tests described in this article a single chip card (SCC) with a FE-I4~\cite{FE-I4} pixel module is used (Figure~\ref{fig:SCC} a). The module is assembled using a high resistive n$^+$-in-n 200 µm thick silicon pixel sensor of the IBL (Insertable B-Layer~\cite{d}) planar design produced by CiS\footnote[1]{CiS, Forschungsinstitut für Mikrosensorik und Photovoltaik GmbH

\quad \url{https://www.cismst.de/en/competences/technologies/}}. Typically, the sensor production process involves aluminum layer deposition on a backside to create an electrical contact to provide sensor biasing (Figure~\ref{fig:SCC} c). Since the infrared light is reflected by that Al layer by 95\%, rectangular openings of 80 $\times$ 30 µm$^2$ were implemented in the backside metalization on top of each 250 $\times$ 50 µm$^2$ pixel for laser illumination (Figure~\ref{fig:SCC} b).

\subsection{The laser test bench setup}
\begin{figure}[htbp]
\centering 
\includegraphics[width=.82\textwidth,origin=c]{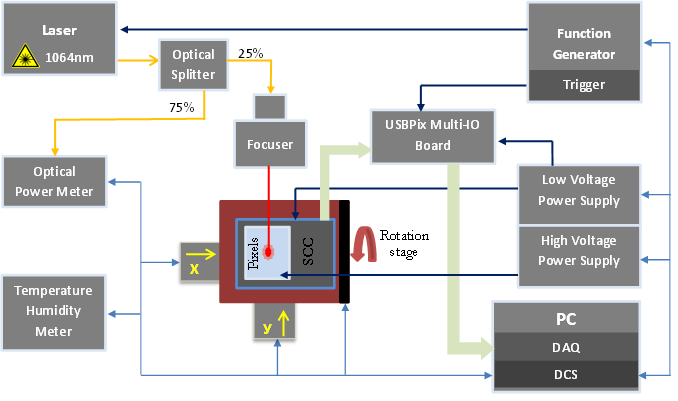}
\caption{\label{fig:scheme}A schematic illustration of the developed laser test setup\cite{thesis}.}
\end{figure}

In this setup, a schematic diagram of which is shown in Figure~\ref{fig:scheme}, a single mode Fabry Perot laser based on Gallium Arsenide (GaAs) diode with a center wavelength of 1064 nm is used to illuminate pixel detectors. The laser diode can operate in a continuous or pulsed mode with 0.5 ns rise time. The pulse power and the rate are controlled by an external excitation signal provided by a function generator. The latter is synchronized with the pixel detector readout electronics. The generated laser light is guided by a single mode optical fiber, and split into two branches with an optical power ratio of 25:75. The least energetic component is then directed to the pixel module surface while the rest is used to monitor laser output power via an optical power meter based on a silicon photodiode.

The device under test (DUT) is mounted on a rotation mechanical stage fixed on a high precision motorized X-Y table from PI\footnote[1]{PI, Physik Instrumente GmbH \& Co. KG

\quad \url{https://www.physikinstrumente.com/en/}} with a minimum motion step of 0.1 µm. The rotation stage allows for variation of the incidence angle simulating inclined tracks. The sensor of the DUT is biased using a Keithley 2410 high voltage power supply via the LEMO cable. The front-end chip is powered by a Hameg HMP4040 low voltage power supply connected by the Molex 8 cable to the SCC.
The mechanical part holding the SCC and optical focusing system are enclosed in a Faraday cage shielding the system from E/M interference and providing a dark and safe operating environment. A dark, light absorbing coating has been applied to all inner surfaces of the enclosure. Optical and electrical pass-throughs are interfaced via a wall mounted patch panel.

The data readout from the module occurs via the RJ-45 cable connected to the USBPix\cite{usbpix} readout system. Communication between the USBPix board and user is established via the STControl\cite{st} software running on a standard PC. The dedicated LabVIEW\texttrademark{} based software has been developed in order to control the devices presented in the system via GPIB and USB interfaces and to simplify the process of position adjustment and overall operation of the setup.

\subsubsection{Mechanical design}
\begin{figure}[htbp]
\centering 
\includegraphics[width=.75\textwidth,origin=c]{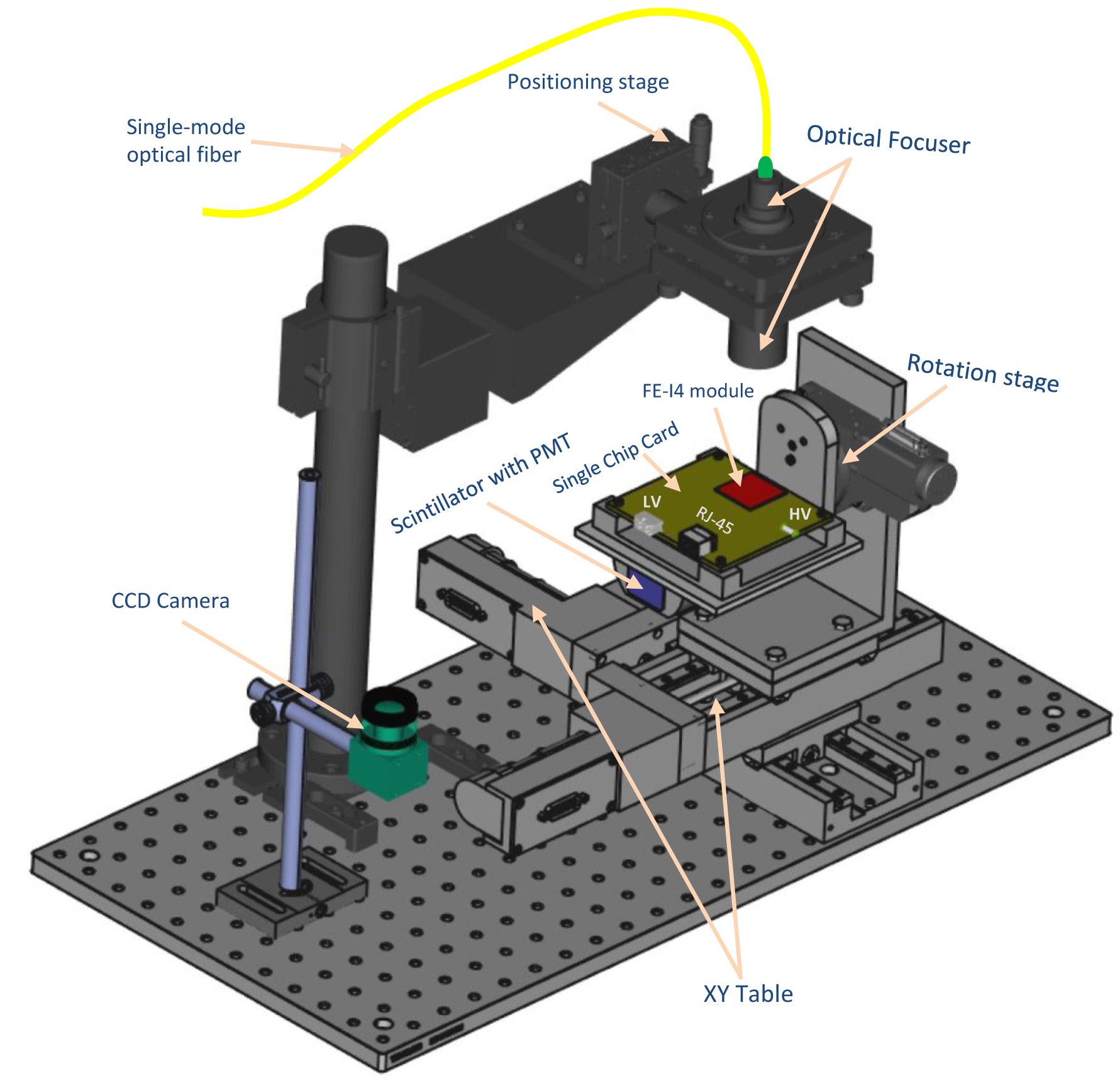}
\caption{\label{fig:laserSet}A 3D model of the laser test bench. The SCC holding a pixel module is attached to the rotation stage, mounted in turn on the X-Y table. The large post holds a focuser. The CCD camera is fixed at the same level as a pixel module for the beam characterization.}
\end{figure}

Special attention is paid to the development of the mechanical part of the setup in order to achieve reliability, repeatability and precision in laser scan measurements. The setup is built on an aluminum breadboard of the 30 $\times$ 70 cm$^2$ size and with M4 threaded holes. The mechanical design is shown in Figure~\ref{fig:laserSet}, with a CCD camera for laser beam measurements.

The X-Y table carries an L-shaped aluminum support used to hold up the rotation stage. The SCC is attached to the rotation stage via a 3D printed support,which is designed to align the plane of the module on the SCC to the rotation axis. It also holds a scintillator coupled to a photomultiplier (PMT) underneath the module, used to provide a trigger signal during measurements with a radioactive source.
The flexibility of such a mechanical construction aims to be used with the focusing systems with different working distances. With the addition of an aluminum support, a quick and easy installation of a radioactive source is also possible. The position of the source can be aligned with respect to the pixel module using a red light laser.

\subsubsection{Focusing system}
An optical focusing system with the large working distance (WD) of about 10 cm is used. The WD is one of the crucial parameters, which defines the design of the focusing system. It is equal to the spacing between the lens and the pixel sensor surface. A large WD is necessary to prevent any damage of both optical lenses and the pixel detector especially, when module rotation is foreseen.
The resulted beam spot size measured with a CCD camera is around 10 µm, wherein the 80 \% of the beam intensity is concentrated in a 2 µm size spot.

\section{The laser test results}
The laser beam has to be aligned, which means that the beam must penetrate the silicon sensor perpendicularly, and well focused before starting the measurements. The preliminary alignment of the system is done using a laser level. A CCD camera is used to verify and refine the beam alignment.
An initial focusing is performed using the CCD camera fixed at the same level as a pixel module. Then the base of the focuser is moved from the position over the camera to the position over the pixel sensor and iterative measurements on the pixel detector are performed. By changing the focusing distance using the manual micrometric stage, one can attempt to achieve the smallest possible beam spot.

\subsection{Laser scans}
\begin{figure}[htbp]
\centering 
\includegraphics[width=.87\textwidth,origin=c]{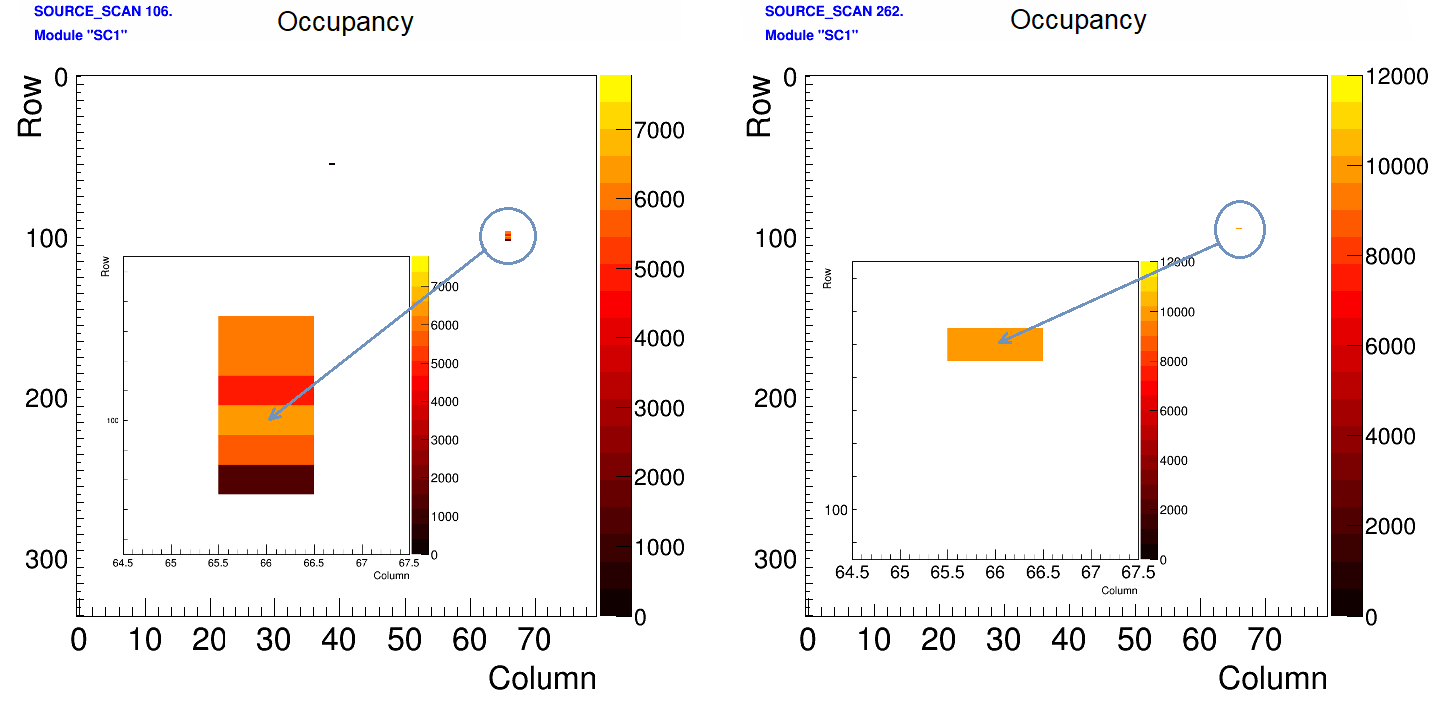}
\caption{\label{fig:hitmaps}The laser scan hitmaps represent two cases: the laser beam spot is seen in several pixels (Left) and in only one pixel (Right). The whole (80 columns and 336 rows) sensor matrix is shown, as well as a zoomed image of the region with hitted pixel channels. The color scale shows the occupancy.}
\end{figure}

To make measurements with the pixel detector illuminated by a laser beam, a standard procedure included in STControl software, called “Source scan” is used. The laser driving signal provided by the external pulse generator is injected as trigger to the USBPix board input. Upon trigger reception, the event information is readout from the pixel module and eventually written into a raw data file for further analysis.
The data file contains information on hit events and their position on the pixel matrix. The data can be plotted in a two dimensional plot containing the number of registered events for each pixel, so-called hitmap. A typical hitmap measured with the laser is shown in Figure~\ref{fig:hitmaps}. 

The Gaussian profile of the beam can extend over several pixels. If the intensity is increased enough, then photons at the halo of the profile may induce sufficient charge in neighboring pixels to exceed the set threshold. Therefore, two cases are considered, namely when only one pixel fired and several pixels fired in a single event. Depending on laser power, a cluster of several pixels that register hits can be obtained. Charge sharing between adjacent pixels and reflection from front side metallization may also contribute in increasing cluster size.

Time of arrival (ToA) information is provided in form of a 16-count, spaced by 25 ns, window, corresponding to consecutive Level-1 on-chip triggers. ToA and time over threshold (ToT) values are recorded in the raw output files. The ToT is the time (measured in clock cycles of 25 ns) that the analog signal induced by the particle has spent over the threshold. It is proportional to the collected charge in the pixel cell. The proportionality is not linear due to secondary order effects like time-walk and non-constant discharge rate~\cite{ToT}. To estimate the relationship between charge and ToT values, a special charge to ToT calibration scan included in STControl software has to be performed. Once this scan is completed, one can extract parameters describing a second order polynomial, which gives the correspondence between the collected in a certain pixel charge $Q_c$ and its time over threshold value $ToT$:
\begin{equation}
Q_c=Par{\_}A+Par{\_}B\cdot ToT+Par{\_}C\cdot ToT^2
\end{equation}
where $Par{\_}A$, $Par{\_}B$ and $Par{\_}C$ are the fit parameters for each pixel obtained from the calibration scan.

Examples of ToA and ToT distributions corresponding to the hitmaps for both cases discussed in Figure~\ref{fig:hitmaps} are presented below in Figure \ref{fig:TotSingle} and Figure \ref{fig:TotSeveral}.
\begin{figure}[htbp]
\centering 
\includegraphics[width=.85\textwidth,origin=c]{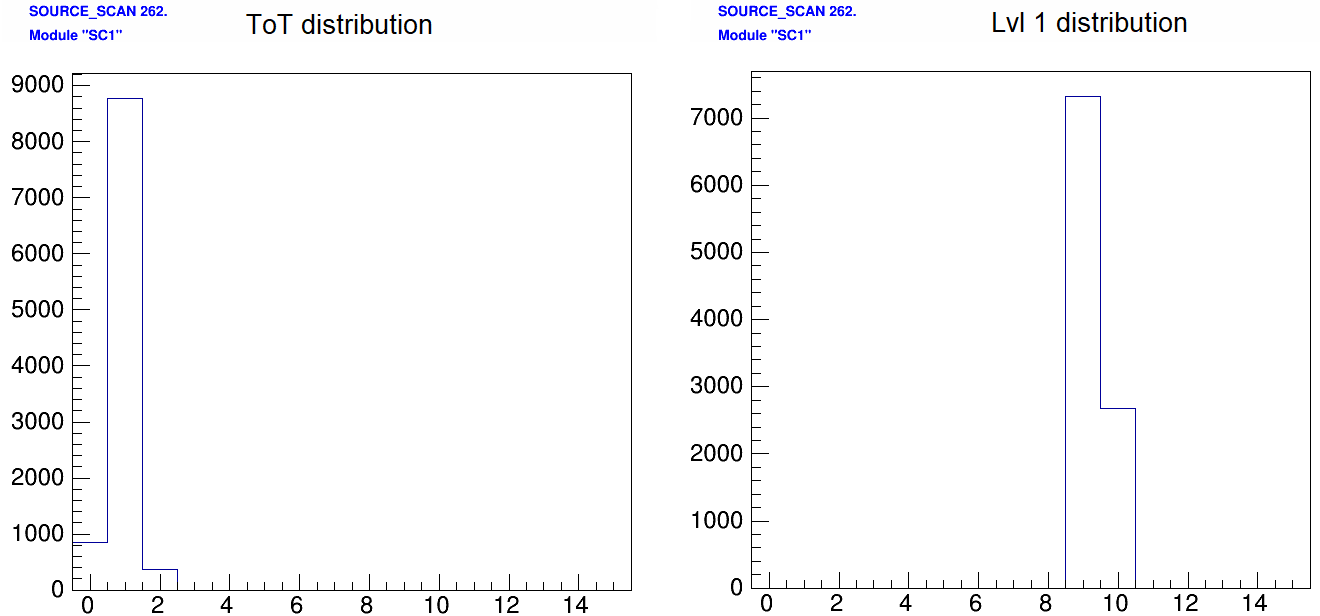}
\caption{\label{fig:TotSingle}ToT (Left) and ToA (Right) distributions for the case where laser generated charge is collected by only one pixel.}
\end{figure}

\begin{figure}[htbp]
\centering 
\includegraphics[width=.78\textwidth,origin=c]{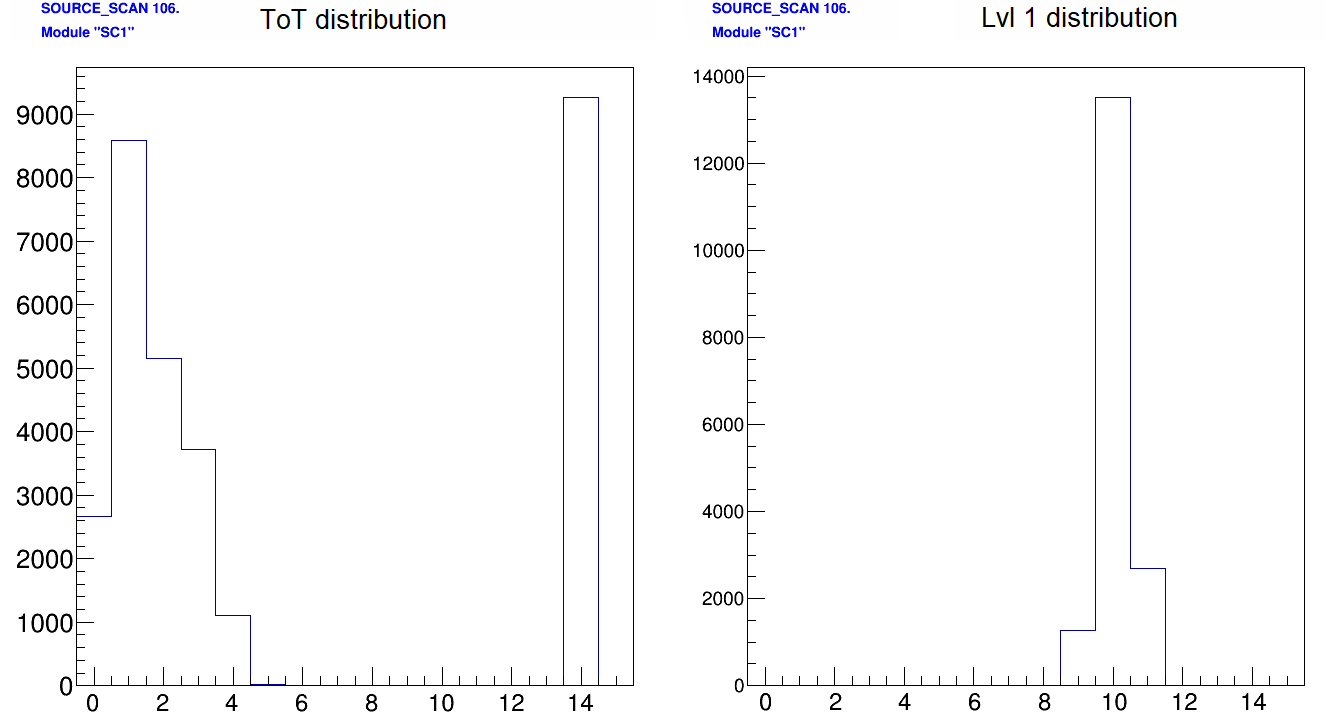}
\caption{\label{fig:TotSeveral}ToT (Left) and ToA (Right) distributions for the case where laser generated charge is collected by several pixels.}
\end{figure}
A ToT distribution for the case, where generated charge is collected by one pixel, takes the form of a peak, with most hits contributing to one bin. Such behavior could be explained by a charge generation mainly through photoelectric effect, so the full energy of photons is transferred to Silicon atoms. The Compton Effect is minor in this energy range. When the laser beam is registered in several pixels, the total generated charge is shared between neighbor pixels, so it results in a much wider ToT distribution.

The ToA distribution for the both cases has a peak shape, demonstrating that most hits have a strong correlation with the timing of the laser pulse signal. In the case when charge is shared between adjacent pixels the ToA distribution may contain a fraction of a late hits due to charge diffusion.

\subsection{Depletion voltage measurements}
The full depletion voltage is one of the most important parameter characterizing silicon pixel sensors. The IV curve is not always an adequate mean to determine full depletion voltage and unfortunately the CV method could not be applicable to the module due to the large addition of capacitance of the components. The laser test bench can function as an alternative method determining the full depletion voltage of the pixel sensor.

\begin{figure}[htbp]
\centering 
\includegraphics[width=.6\textwidth,origin=c]{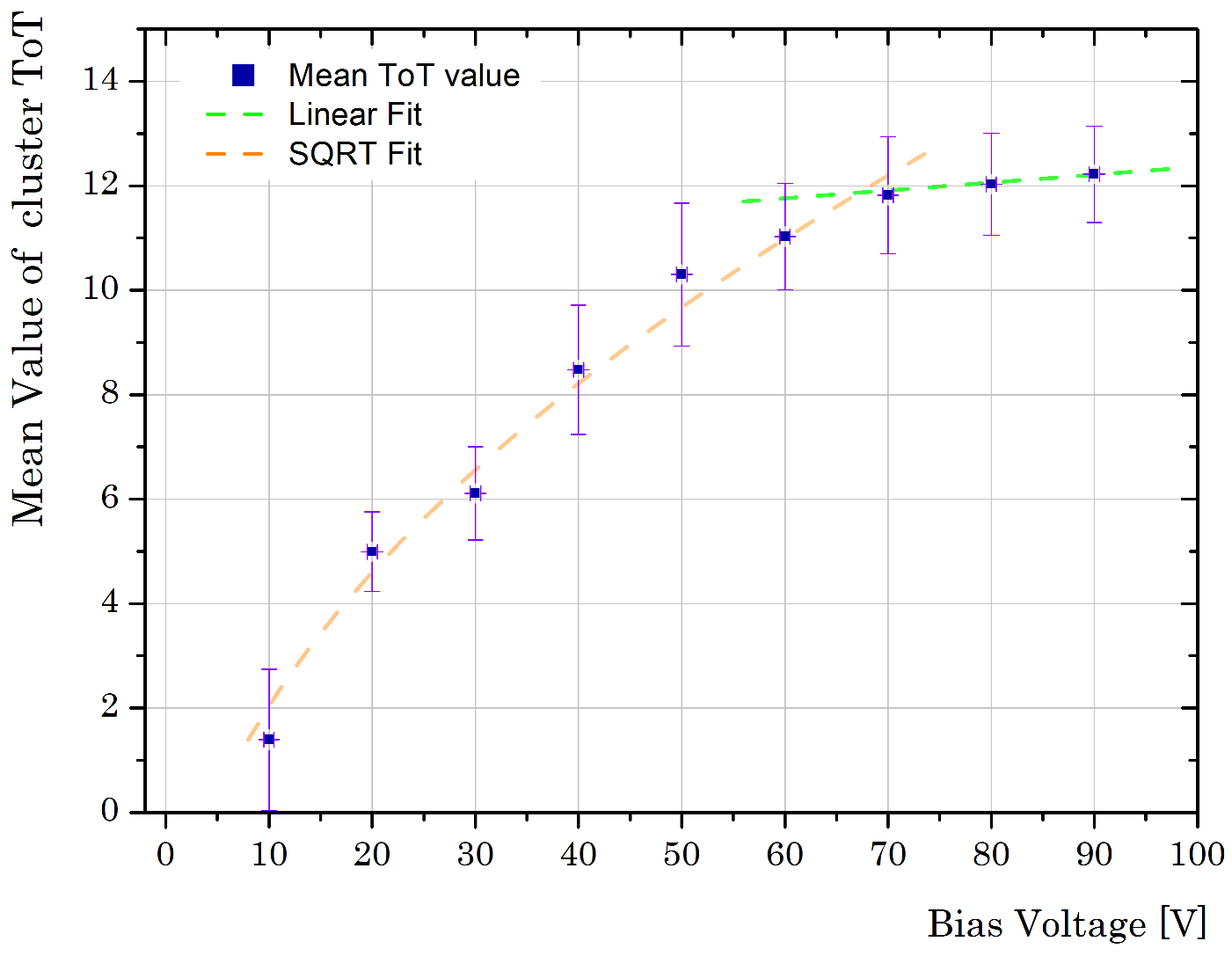}
\caption{\label{fig:Depletion}The charge signal amplitude of the seed pixel versus the applied bias voltage. The collected charge increases with $\sqrt{V}$. At full depletion the collected charge shows saturation.}
\end{figure}

Plotting, as shown in Figure \ref{fig:Depletion}, the mean ToT value of the hits registered during the laser illumination, versus bias voltage, one can find the turning point, when the dependence reaches a plateau in mean ToT value. This point is interpreted as a full depletion voltage. Employing this method a depletion voltage of the sensor of IBL type, used in the test, has been found to be about 67 V, which is inside the range of 50 - 70 V \cite{d} expected for these sensors.

\section{Pixel module characterization using a radioactive source}

Two types of radioactive sources can be used for pixel module characterization. The first type is $\gamma$-sources, which emit $\gamma$-quants with well-known energy spectrum. These $\gamma$-sources are utilized as the collected charge reference. The second type is $\beta$-sources, which emit electrons, and can be used for charge collection efficiency study.

\begin{figure}[htbp]
\centering 
\includegraphics[width=.9\textwidth,origin=c]{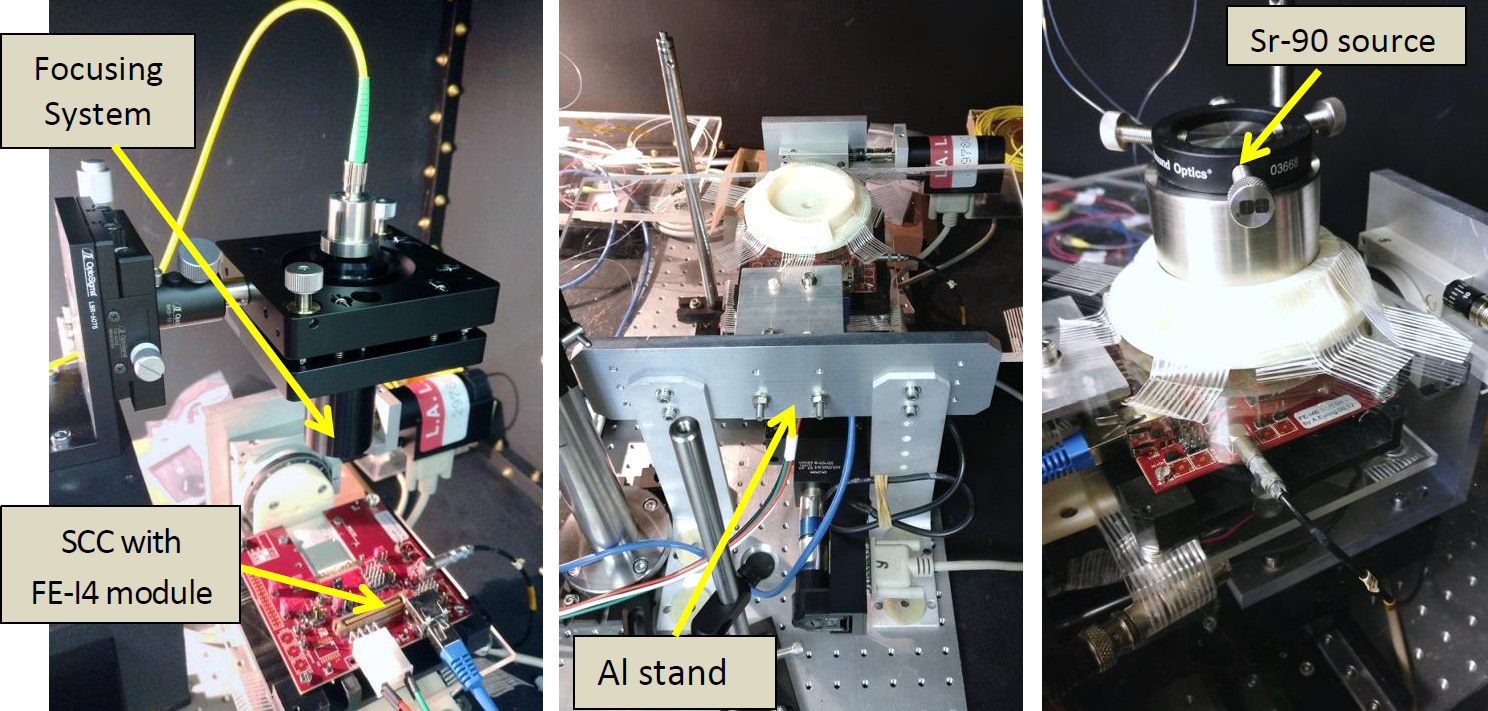}
\caption{\label{fig:SourceTest}A picture of the setup to perform the laser tests (Left). The aluminum stand is add-on installed to the laser setup to support radioactive source for the source measurements (Middle). A picture of the module, placed in between a source and a scintillator during the measurements (Right).}
\end{figure}

The developed laser test bench setup may be easily transformed to a radioactive source or cosmic rays configuration. A source is placed on an add-on aluminum stand, shown in Figure~\ref{fig:SourceTest}, with a polycarbonate plate and the 3D printed support for the positioning of the source over the pixel sensor. To facilitate operation the laser beam holder can be removed. 

A 90-Sr source of nominal activity 37 MBq, encapsulated in a stainless steel support with a circular window, has been used. Isotope of Strontium undergoes $\beta$-decay to Yttrium, which in turns emits a beta spectrum of electrons with the maximum of 2.28 MeV energy:
\begin{equation*}
\label{eq:x}
\begin{split}
^{90} Sr\, \rightarrow \, e^- +\overline\nu _e +\: ^{90}Y \,,
\qquad
t_{half}=28.8\,y \,,\quad E_{max}=0.55\, MeV\,,
\\
^{90} Y\, \rightarrow \, e^- +\overline\nu _e +\: ^{90}Zr\,,
\qquad
t_{half}=2.67\,d \,,\quad E_{max}=2.28\, MeV\,.
\end{split}
\end{equation*}

In a radioactive source configuration, two ways are possible for the pixel detector readout triggering. The first option is to use an external trigger signal from a scintillator fixed below the detector. The scintillator output signal can be shaped with an external discriminator, so that only hits above a certain level cause a trigger signal. Thus, we can reduce noise and suppress the low energy electrons from a beta spectrum, thereby ensure the energy deposition caused by only high energy particles. Also an arbitrary clock cycle can be applied to the external trigger input for acquiring the data. In this mode (commonly named random trigger) all the hits within the clock interval are read out.

A second option is so-called self-triggered mode. For this purpose the internal trigger of the readout chip is used. Once the charge is detected, the trigger signal is generated to activate readout process. This option is mostly used for the $\gamma$-sources, where the photons release all their energy in a single shot when they interact with the atoms in silicon bulk, so cannot reach the scintillator.

\begin{figure}[htbp]
\centering 
\includegraphics[width=.9\textwidth,origin=c]{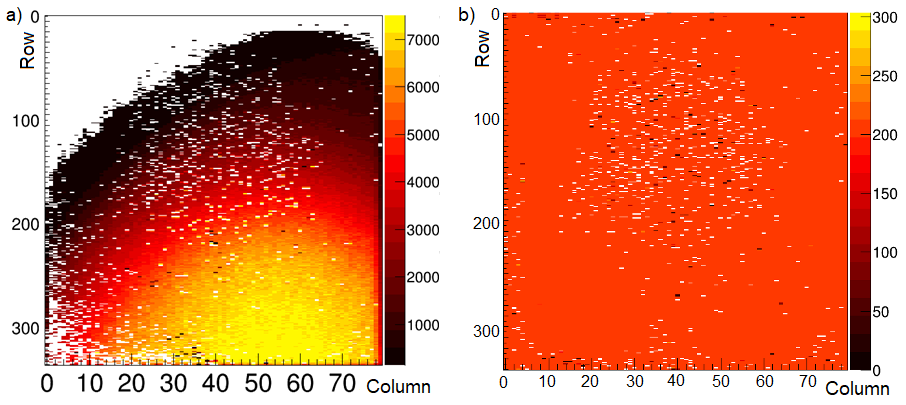}
\caption{\label{fig:SourceResult}a) The resulting 2D hitmap of the source scan. The color scale shows the occupancy. Area with more events corresponds to the source position, which is over the bottom right corner of the sensor. The FE-I4 type module was non-irradiated and tuned to a threshold of 2500 e$^-$ for the test at room temperature. The noisy and stuck pixels were masked. The device was biased to -70 V. \hspace{2 mm} b) A 2D analog scan hitmap shows how many times each pixel has responded, the charge was injected 200 times for every pixel. One can see the noisy pixels in yellow, black pixels with a very high threshold and disabled pixels in white.}
\end{figure}

The resulting hitmap taken during a radioactive source scan with an external random trigger is presented in Figure~\ref{fig:SourceResult} a. The source measurements allow to check the pixel detector functionality starting from charge collection in the sensor, charge transfer to the front-end chip, where the signal is processed, to data readout. For example, comparing the hitmap from the source scan (Figure~\ref{fig:SourceResult} a) with the chip analog scan (Figure~\ref{fig:SourceResult} b), one can see a group of “white” pixels with disconnected bump bonds in left bottom and right up corner.

\section{Conclusion}
The test bench setup based on an infrared laser for silicon pixel detectors characterization has been developed and tested. The software is written to optimize the preparation of the setup for the tests and control it during the measurements. A small beam spot size and accurate detector positioning make it possible in future to make in-pixel efficiency scans over the whole sensor sensitive surface. The data can be quickly analyzed as no track reconstruction is needed. The setup is versatile with minimal effort, it can be adapted for the measurements with a radioactive source. This system is extremely useful for quality control of the modules to build silicon trackers of modern and future high-luminosity collider experiments, especially for pixel module production for the HL-LHC. It is a quick, still precise and affordable option, comparing to X-ray imaging technique, for checking the bump bonding connection to find merged or disconnected bumps. The feasibility of such a system for the testing of pixel modules has been demonstrated in this paper. 

Such a table-top laser test bench setup would be of a great help in pixel and strip detector characterization, complementary to the particle beam tests. In the foreseeable future this laser test bench will be equipped with a new DAQ system allowing to readout ITk pixel modules. Furthermore the setup will be integrated and coupled to instruments such as probe station to further enhance the laboratory in-house test capabilities for the pixel detectors.


\acknowledgments

The authors would like to acknowledge Maurice GLASER from CERN for providing of the semiconductor infrared laser and Julian DAJCZGEWAND from Optosigma Ltd. for the support in production of the focusing system, making this work possible.


\end{document}